\pdfoutput=1

\documentclass[aps, prl, twocolumn]{revtex4-2}

\usepackage{amsmath}
\usepackage{amssymb}
\usepackage{color}

\usepackage{hyperref}

\def\flushboth{%
  \let\\\@normalcr
  \@rightskip\z@skip \rightskip\@rightskip
  \leftskip\z@skip
  \parindent 1.5em\relax}

\usepackage{verbatim}
\usepackage{graphicx}
\usepackage{subfig}

\newcommand{\bi}{\begin{itemize}}\newcommand{\ei}{\end{itemize}}
\newcommand{\be}{\begin{equation}}\newcommand{\ee}{\end{equation}}

\newcommand{\bml}{\begin{multline}}
\newcommand{\emll}{\end{multline}}

\newcommand{\nn}{\nonumber}

\def\({\left(} \def\){\right)}\def\[{\left[} \def\]{\right]}\def\Re{\text{Re}}\def\Im{\text{Im}}

\def\ho{\hat \omega}\def\hn{\hat n}

\def\al{\alpha}
\def\mO{\mathcal{O}}
\def\eps{\epsilon}
\def\v{\vec}

\def\g{\gamma}\def\lam{\lambda}

\def\d{\partial}

\newcommand{\la}{\langle}\newcommand{\ra}{\rangle}
\newcommand{\bea}{\begin{eqnarray}}\newcommand{\eea}{\end{eqnarray}}

\def\ie{\begin{equation}\begin{aligned}}
\def\fe{\end{aligned}\end{equation}}

\def\tilde{\widetilde}\def\hat{\widehat}

\def\d{\partial}\def\1{{\mathds 1}}\def\Im{\mathop{\rm Im}}

\def\mL{\mathcal{L}}

\def\o{\omega}

\begin{document}

\title{Interaction renormalization and validity of kinetic equations for turbulent states}

\author{Vladimir Rosenhaus}
\affiliation{Initiative for the Theoretical Sciences, Graduate Center\\ CUNY,
365 Fifth Ave, 
New York, NY 10016}
\author{Gregory Falkovich}
\affiliation{Weizmann Institute of Science,  Rehovot 7610001 Israel}


\begin{abstract}

We consider turbulence of waves that interact weakly via four-wave scattering (sea waves, plasma waves, spin waves, and many others). In the first non-vanishing order in the interaction, the occupation number of waves  satisfy a closed kinetic equation which has  stationary solutions describing turbulent cascades. We show that a straightforward perturbation theory beyond the kinetic equation gives terms that generally diverge both at small (IR) and large (UV) wavenumbers for a direct cascade. The analysis up to the third order identifies the most UV-divergent terms. In order to gain qualitative analytic control, we sum a subset of the most UV divergent term, to all orders, giving a perturbation theory which is generally free from UV divergence, showing that turbulence becomes independent of the dissipation scale when it goes to zero.  On the contrary, the ever-present IR divergence means that the effective coupling is parametrically larger than the naive estimate and grows with the pumping scale  (similar to  anomalous scaling in fluid turbulence). {\it This suggests that the kinetic equation does not describe wave turbulence even of arbitrarily small level if the cascade is sufficiently long.} We show that the character of strong turbulence is determined by the sign of the coupling, that is,  whether the effective four-wave interaction is enhanced or suppressed by collective effects. The enhancement possibly signals that strong turbulence is dominated by multi-wave bound states (solitons, shocks, cusps), similar to confinement in quantum chromodynamics. 
\end{abstract}
\maketitle

\maketitle

\section{Introduction}

 Kinetic equations are workhorses of physics and engineering. They describe transport phenomena and turbulence, from gas pipes and oceans to plasmas in space and in thermonuclear reactors. These equations are so effective because they reduce the description of multi-particle or multi-wave systems to a closed equation on a single-particle probability distribution or a 
single-wavevector occupation number. The equations have solutions that describe thermal equilibria, transport in weakly non-equilibrium states, and even far-from-equilibrium turbulent states. The question is whether these solutions are physical, that is, if  accounting for multi-particle and multi-mode correlations leads only to small modifications.

This question was first addressed for the density expansion beyond the Boltzmann equation describing binary collisions \cite{Bog,Dorfman,DC2,DC3,DC4,KO}. 
The higher-order terms involve subsequent collisions of the same particles and contain spatial integrals over the positions of intermediate collisions. These integrals have infrared (IR) divergences starting from the second order (in 2D) or from the third order (in 3D). The  divergences reflect the memory effects creating long-distance multi-particle correlations. In thermal equilibrium, such divergences are canceled due to detailed balance, and the correlations are short so that the equations of state have a regular virial expansion. Spatial non-equilibrium prevents cancelation in transport states. Of course, the divergences appear because the ``naive" virial expansion allows particles to travel arbitrarily long distances between collisions. One must account for the collective effects that impose the mean free path as an IR cutoff.  The renormalized expansion then involves powers of density other than integers (adding density logarithms in this case). Such perturbation theory is singular, even though the corrections are small in dimensions exceeding two. The divergences lead to logarithmic renormalization of the kinetic coefficients in two dimensions and to anomalous kinetics in one dimension, which is a subject of active research.

In contrast to transport states, turbulent states create fluxes (cascades) in momentum space rather than in real space. The cascade distributions were found as exact (Zakharov) solutions of the kinetic equations both for particles and waves assuming locality of interactions, which is the convergence of the collision integrals \cite{ZLF}. This means that the contribution of the lowest-order collisions and interactions is predominantly given by comparable momenta of colliding particles or wavenumbers of interacting waves.  The  question is then whether locality also holds in the higher-order corrections \cite{RS1,RS2}, which describe multi-particle collisions and multi-wave interactions. In this work, we answer this question in the negative, finding  in $k$-space (where non-equilibrium now lives) the divergences that bring a new four-wave coupling renormalized by multi-wave interactions.

We consider  particles/quasiparticles interacting via four-wave scattering described by the Hamiltonian
 \be \label{H21}
H = \sum_p \o_p |a_p|^2 + \sum_{p_1, p_2, p_3, p_4}\!\! \lambda_{1234} a_{p_1}^*  a_{p_2}^* a_{p_3} a_{p_4}~.
\ee
The bare values of the   frequency $\omega_k$ and the four-wave coupling $\lambda_{1234}$  are given by the equations of motion which are uniformly valid on all scales, including at and below our UV cutoff, which is the dissipation scale of turbulence. A single wave propagating in an undisturbed medium has the frequency $\omega_k$, and  $\lambda_{1234}$ defines the interaction energy of four waves without any other waves present. Both  $\omega_k$ and   $\lambda_{1234}$  are renormalized in a multi-mode turbulence state, which is the subject of this work. It is important to stress that the nature of renormalization in turbulence theory is quite different from that in quantum field theory or critical phenomena. There, one always deals with effective large-scale theories, describing how a (generally unknown) bare value at some small UV scale (Planck scale or lattice spacing) is getting renormalized as one increases the scale of measurements.   In quantum field theory, one cannot switch off quantum fluctuations, so the meaning of  renormalizion is different.

We wish to find how the renormalization depends on the turbulence level and the extent of scales and how this dependence changes the weakly turbulent Zakharov spectra, if we keep the turbulence level small at a given $k$, increasing either the pumping scale or the dissipation wavenumber. In particular, we show that the ever-present IR divergences in the renormalization of $\lambda_{1234}$ make deviations from  weak turbulence grow with an increase of the pumping scale $L$ (for a direct cascade).  This introduces an explicit dependence on $L$, reminiscent of an anomalous scaling in fluid turbulence,  where the statistics at a given wavenumber $k$ deviates further and further from Gaussian as $kL$ increases. This suggests that a sufficiently long cascade interval $kL$ makes wave turbulence strong and the Peierls kinetic equation invalid, even at a small level of nonlinearity at a given $k$.

{\bf The model}. We assume  $\o_k = k^{\al}$ with $0<\alpha\leq2$. The real vertex $\lam_{1234}  \equiv \lam_{p_1 p_2 p_3 p_4}\delta({\bf p_1}{+}{\bf p_2}{-}{\bf p_3}{-}{\bf p_4}) $ can be of either sign and is a homogeneous function of degree $\beta$: $ \lam_{ap_1 ap_2 ap_3 ap_4}=a^\beta  \lam_{p_1 p_2 p_3 p_4}$. 
The complex amplitudes $a_k$ satisfy the equations of motion $ida_k/dt=\partial H/\partial a_k^*$. 
The time derivative of the occupation numbers $n_k=\langle |a_k|^2\rangle$ is expressed via the fourth moment,
\be \label{424v2}
 {\partial n_k\over \partial t}     =I_k= 4  \sum_{p_2, \ldots, p_4}  \lam_{k234}\text{Im} \la a_{k}^*  a_{p_2}^* a_{p_3} a_{p_4}\ra ~.
\ee
The equal-time fourth moment in (\ref{424v2})  can be found perturbatively as a series in $\lambda$, assuming the statistics to be close to the Gaussian statistics of non-correlated waves, ${\cal P}\{a_k\}\propto \exp[-\sum_k|a_k|^2/n_k]$, which is completely determined by the occupation numbers. This provides the zeroth-order approximation, where the right-hand side of (\ref{424v2}) is zero. The standard perturbation theory is described in the Appendix. It gives in the first non-vanishing order the Peierls kinetic equation (KE):

 \bea 
{\partial n_k\over \partial t}  =I_k=&16\pi  \!\!\sum\limits_{p_2, p_3, p_4} \!\! \lam_{k234}^2\,n_k n_2n_3n_4
 \nonumber\\&\times
\delta(\o_{k2;34})\Big( \frac{1}{n_k} {+} \frac{1}{n_2}{-}\frac{1}{n_3} {-} \frac{1}{n_4} \Big)   ~,\label{32}
\eea
where we defined $ \o_{k2;34} \equiv\o_{k} {+} \o_{p_2}{-}\o_{p_3}{-}\o_{p_4}$.  
What is traditionally required for the validity of \eqref{32} is to provide a dense enough set of such resonances, which requires taking the limit  $kL\to\infty$ (apart from $\lambda_{1234}\to0$), where $L$ is the box size, see e.g. \cite{Math1,Math2,Math3}. This fact already requires a careful analysis of divergences in the kinetic equation and in the corrections to it.

The leading-order kinetic equation (\ref{32}) conserves energy $\int \omega_kn_k\,d\v k$ and  wave action  $\int n_k\,d\v k$, and has two stationary solutions which describe turbulent cascades. Here we focus on the direct energy cascade (from small to large wavenumbers). Writing  (\ref{32}) as the energy continuity equation, $k^{d-1}\omega_k{\partial n_k/ \partial t} =-\partial P_k/\partial \v k$, and requiring the spectral flux $P_k=P$ to be $k$-independent,, we obtain by power counting 
\be
P_k=k^{d}\omega_kI_k\propto k^{2\beta+3d}n_k^3\ \Rightarrow\ n_k=k^{-d-2\beta/3}~.\label{gp}
\ee
The flux value is chosen to get a factor of unity in front. One can also obtain (\ref{gp}) by estimating the flux $P$ as the energy density $\omega_k n_k k^d$ divided by the nonlinear interaction time $1/\omega_k\epsilon_k^2$, where 
 the dimensionless parameter of nonlinearity (coupling) at a given $k$ can be estimated taking for the sake of power counting $ \lam_{kkkk}\simeq \lambda k^{\beta}$:\be \label{39}
\epsilon_k =  \frac{\lam_{k} n_k k^{d}}{\o_k}\simeq \lambda k^{\beta/3-\al}\!\!=\epsilon_0\left({k\over k_0}\right)^{\beta/3-\al}\!\!\!.
\ee
The standard claim is that (\ref{32}) is valid and (\ref{gp}) is its solution for those $k$ for which $\epsilon_k\ll 1$ and under the conditions of locality, which means convergence of the integrals in (\ref{32}) upon substituting (\ref{gp})  \cite{KK,ZLF,Naz}). That depends on the three asymptotics when one or two wavenumbers become much smaller than the others:
\bea &\lim_{p_1,p_3\ll p_2,p_4}&\lam_{1234}=\lambda (p_2p_4)^{\beta_1/2}(p_1p_3)^{(\beta-\beta_1)/2}\,,\label{beta_1}\\&\lim_{p_1,p_2\ll p_3,p_4}&\lam_{1234}=\lambda (p_3p_4)^{\beta_2/2}(p_1p_2)^{(\beta-\beta_2)/2}\,,\label{beta_11}\\&
\lim_{p_1,p_3,p_4\gg p_4}&\lam_{1234}=\lambda_{12,1+2}  p_4^{\beta_3}\,.\label{beta_2}\eea
For most cases,  $\beta_1=\beta_2=\beta/2$.  For spin waves,   $2\beta_1=\beta_2=\beta=2$, $\alpha=2$ so that $\lim_{p_2\to0}(\v p_3 {\cdot} \v p_4)\propto p_2$ and $\beta_3=1$. In some cases, $\beta_1\leq\beta/2$ provides stronger convergence than might be expected on dimensional grounds, as for surface water waves: $\beta=3$, $\beta_1=1$, $\beta_3=3/4$.

Consider  (\ref{32}) when ${ p}_2 ,p_4\rightarrow \infty$ while $ p_3$ remains finite.
For  $ n_k=k^{-\g}$, the powers of $p_2$ in the terms in  (\ref{32}) are as follows: 
$ | \lam_{1234}|^2 \sim p_2^{2\beta_1}$, $
  \prod_{i=1}^4 n_i \sim p_2^{-2\gamma}$. 
Expanding   $ \o_4= |{\bf p}_1{ +} {\bf p}_2 {-} {\bf p}_3|^{\al}$ at large $p_2$, we find that $\o_2 - \o_4\propto p_2^{\al-1}$. For $\al>1$, this means that $\o_{12;34}\sim \o_{2} - \o_4$, and  $ \delta(\o_{1 2;3 4}) \sim p_2^{1-\al}$. For $\al<1$, we have $\o_{12;34} \sim \o_{1}{-}\o_{3}$, and $ \delta(\o_{1 2;3 4}) \propto p_2^0$. We denote $\kappa= \max\{0,\alpha-1\}\geq 0$, then $\lim_{p_2,p_4\gg p_1,p_3} \o_{1 2;3 4}\sim p_2^\kappa$. Using ${n_4}^{-1}  =\o_4^{\frac{\g}{\al}} = (\o_1 + \o_2 - \o_3)^{\frac{\g}{\al}}~$ we obtain a cancellation  by a factor of $1/\o_2$ in the second line: 
\be \delta(\omega_{k2;34})(n_2^{-1}-n_4^{-1}) \propto  p_2^{\g - \al-\kappa}\,\label{lim}\ee 
Combining all the pieces and setting $\gamma=d+2\beta/3$ gives the convergence in the UV if  $2\beta_1-2\beta/3-\alpha-\kappa<0$.

IR-convergence also depends on  $\al-1$. If $\al>1$, then a 3-wave resonance is possible, and the main IR  contribution is $\int_{k_0} d{\bf p_2}\, p_2^{\beta_3 - \g}~\propto k_0^{\beta_3-\beta/6}$. It converges for spin waves. Optical and plasmon turbulence in a nonisothermal plasma with  $\alpha=2$, $\beta=0$ is IR borderline.
For $\al<1$, no three-wave resonance is possible and we must also take both $p_2,p_3\to0$.
Expanding $ \delta(\o_{1 2;3 4})$ up to  $|{\bf p}_2{-}{\bf p}_3|^2$ we obtain:
\bea
&\int_{k_0} d{\bf p_2}\, p_2^{\beta/2 - \g}\int_{k_0} d{\bf p_3}\, p_3^{\beta/2 - \g}\delta(\omega_{k2;34})\nonumber\\&
\times[(\omega_k{+}\omega_2{-}\omega_3)^{\g/\alpha}-k^\g]\propto k_0^{2\beta/3-2\beta_1-\alpha+2}~.\nonumber
\eea
For $\alpha{<}1$, the combination of the IR and UV conditions gives $2 (\beta_1 {-}\beta/3){+}1{-} 2\al<0<\beta/3{-}\beta_1{+}1{-}\alpha/2$ or $  (\beta_1{ -}\beta/3) {<}  \alpha/2$, which is satisfied for water waves ($\alpha=1/2$).

The form of the kinetic equation thus gives two cancellations in the IR and one in the UV, which provide a locality window for $\gamma$ \cite{ZLF}.
The locality is expected to guarantee that the solution does not depend on $k_0$ (the IR cutoff) and $k_{m}$ (the UV cutoff) in the limits   $k_0/k \to 0$ and $k/k_{m}\to0$. For  $\alpha\leq1$ and $\beta_1=\beta/2$, the convergence condition $\beta/3\leq\alpha$ also guarantees that the nonlinearity parameter $\epsilon_k$ decays with $k$, making it seem as if the weak turbulence approximation only gets better along the cascade. We will see that the locality window is generally absent for higher-order scattering processes, so the validity of weak turbulence needs re-examination. 

When $\beta>3\alpha$ and $\epsilon_k$ grows along the cascade,  on dimensional grounds one might have expected strong turbulence to appear at the $ k$ for which $\epsilon_k\simeq 1$. We will show below that the effective dimensionless coupling parametrically exceeds $\epsilon_k$ so that strong turbulence must start at lower $k$ than had been expected. 

{\bf Next-to-leading order}.  The first correction replaces the bare frequencies in the kinetic equation by the renormalized ones:
$\tilde\o_k=\o_k+\sum_{\bf q}\lambda_{kqkq}n_q$. For optical turbulence, $\beta=0$, $n_q=q^{-d}$, the sum diverges logarithmically; this case will be analyzed elsewhere. For the rest of the cases, the sum always converges in the UV.  For water waves,  $\lambda_{kqkq}\propto q({\vec k\cdot\vec q})$ for $q\ll k$ -- angular integration cancels the IR divergence; it cancels the whole one-loop contribution for spin waves. A power-law IR divergence takes place for plasma turbulence with $\beta=2=\alpha$ but the frequency renormalization grows slower with $k$. Even when the one-loop frequency renormalization is comparable to $\omega_k$, the  replacement $\omega_k\to\tilde\omega_k$ does not change the energy cascade spectrum. Higher-order corrections to the Green functions contain real and imaginary parts; they have increasingly higher powers of divergence, which will be addressed elsewhere. 
Here we conclude that the lowest-order frequency renormalization does not bring about substantial changes in the weak-turbulence spectra.

The next-to-leading order renormalization of the quartic interaction gives the contribution to $\partial n_k/ \partial t$ which is KE \eqref{32} multiplied by the sum of two loop integrals $\mL_a+\mL_b$ \cite{RS1, RS2}:
\bea &\mL_a=4 \sum\nolimits_{p_5,p_6}(n_5+n_6)
    { \lam_{k256}\lam_{5634}  }}/{ \lam_{k234}\o_{k256}\,,\label{aterm}\\
&\mL_b=16\sum\nolimits_{p_5,p_6} (n_5-n_6)  {\lam_{k645}\lam_{2536}  }/{\lam_{k234} \o_{3625}}\,.\label{bterm}
\eea
The integrals of $1/\o$ are the principal value part. As described in the Appendix, a and b correspond to parallel or anti-parallel arrows in the bubble diagram. For thermal equilibrium, $n_k \sim 1/(\o_k{+}\mu)$, the corrections vanish, at all orders. 
This is not so for the turbulent solution. 
Let us now substitute the weak-turbulence spectrum $n_k=k^{-d-2\beta/3}$ into the new equation and compare convergence with that of  (\ref{32}).  There are two new convergence issues here: an extra (loop) integration over $p_5$ and extra powers of $p_2,p_3, p_4$ in the external integration.  The conditions for UV-convergence of the loop integration are the same as for the bare KE (except for spin waves described in the Appendix). Here we assume the loop integration is UV-convergent.  

IR divergence of the loop integration is determined by the limit in which one wavenumber goes to zero, which gives $k_0^{2\beta_3-2\beta/3}$ for any $\alpha$ since $p_5$ is not bound to the resonance surface. Since in all cases (except spin waves)
$\beta_3=\beta/4$,  the divergence scales as $k_0^{-\beta/6}$, including for water waves. Setting in (\ref{aterm},\ref{bterm}) $p_5=k_0$ and ${\bf p_6}={\bf k_1}+{\bf k_2}$, we obtain the addition to the vertex $\lambda_{1234}$ in the following form
\bea \delta\lambda_{1234}& =\lambda_{1234}\left(\mL_a+\mL_b\right)=  {96\Omega_d k_{0}^{-\beta/6}\over\beta  (2\pi)^d }\Bigl({\lambda_{1,2;1+2}\lambda_{3,4;3+4}\over \omega_1+\omega_2-\omega_{1+2}} \nonumber\\&-{2\lambda_{1,4-1;4}\lambda_{2;3,2-3}\over \omega_4+\omega_{4-1}-\omega_{1}}-{2\lambda_{4,1-4;1}\lambda_{3;2,3-2}\over \omega_4+\omega_{1-4}-\omega_{1}}\Bigr) \label{V0}\,.\eea

Let us see how this new effective vertex affects the  UV divergences in the integration over $p_2,p_3,p_4$. When $p_2,p_4, p_6\to\infty$ we have  ${\o_{k2; 56}}  \sim  p_2^{-\kappa}$,  $\lam_{1256} \lam_{5634}/\lam_{1234} \sim p_2^{\beta_1}$, and similarly for the  b-term.  Thus the extra factor relative to (\ref{32}) brings an extra $p_2^{\beta_1 - \kappa}$ into  the integrand.
The power $\beta_1-\kappa$ is non-negative for all cases with $\alpha<1$ and for some cases with $\alpha>1$ (plasma turbulence). 
This power counting suggests that if $\beta_1-\kappa\geq0$, then starting from the  $m$'th order,  where $m$ is such that $\beta/3- \al- \kappa +m(\beta_1 - \kappa)\geq0$, the perturbation theory brings terms whose degrees of UV-divergences  grow linearly with $m$. However, this power counting is incorrect because the leading UV divergences cancel in the one-loop corrections. Indeed, for $p_2,p_4\gg p_1,p_3$, we have $\lambda_{1,2;1+2}\approx\lambda_{1,4-1;4}\approx\lambda (p_1p_2)^{\beta/4}$. The denominator makes the last term in  (\ref{V0}) small, while the symmetry of the first term  brings an extra factor 2 so that   (\ref{V0}) is proportional to $\omega_{2-3}+\omega_{2+1}-\omega_2-\omega_4\approx p_{1i}p_{3j}{\partial^2 \omega_2/\partial p_{2i}\partial p_{2j}}$. As a result, 
\be\!\!\!\!{\delta\lambda_{1234}\over \lambda_{1234}}\propto {\lambda\partial^2 \omega_2\over\partial p_{2i}\partial p_{2j}}{{p_2^{\beta_1} p_{1i}p_{3j}\over[\omega_1-({\bf p}_1 {\bf v}_2)] [\omega_3-({\bf p}_3 {\bf v}_2)]} }\,.\label{UUVV}\end{equation} Here ${\bf v}_2=\partial\omega_2/\partial {\bf k}_2$.
The one-loop correction adds to the kinetic equation  a power of $\beta_1-\kappa+\alpha-2$, instead of  the naive $\beta_1-\kappa$, which means convergence for all cases of interest including surface gravity waves with $2\al = \beta_1=1=\beta/3$.

Yet, the roller-coaster does not stop there. A lengthy computation of the two-loop contributions is presented in the Appendix. All diagrams, as expected, have doubled IR divergence $k_0^{-\beta/3}$. Yet the power of the UV divergences allows us to sort the two-loop diagrams. Each one  adds to the vertex schematically $\lambda_{12ij}\lambda_{ijkl}\lambda_{kl34} (n_i+n_j)(n_k+n_l)/\omega_{12ij}\omega_{kl34}$. When $p_i,p_k\to\infty$ the vertices give the power $2\beta_1$, while every frequency is expected to give $-\kappa$ as in the KE and one-loop above. This is indeed so for the bubble diagrams adding to KE the UV factor $ k_m^{2\beta_1-2\kappa}$. They are essentially the squares of the first two terms from the last bracket in \eqref{V0}, so they add rather than cancel. They couldn't be canceled by the non-bubbles (mixtures of a and b terms) which are subleading due to internal cancellations similar to \eqref{lim}, see (\ref{lim0},\ref{lim1},\ref{lim2}); they add to KE factors like $k_m^{2\beta_1-\alpha-\kappa}$,   $k_m^{2\beta_1-2\alpha}$. That means that UV divergences at higher orders are real and need to be taken care of.

This requires summing the most UV divergent diagrams at each order, a challenging task. We will do something simpler, which  is to sum just that the bubble diagrams (sequential iterations of the one loop diagram). This will be sufficient for curing the UV divergence, but we do not anticipate that this gives a quantitatively correct answer in general. The bubble diagrams are summed  via an integral Schwinger-Dyson equation in the momentum-frequency domain, see \eqref{SDeqn} in the Appendix.  We were able to solve this equation explicitly  for two particular classes of the bare vertex. 
The simplest case in which  one can explicitly sum all the bubble diagrams is when the ratio  
${\lam_{1256}\lam_{5634}}/{\lam_{1234}}$ is only a function of $p_5$ and $p_6$.
The sum of bubble diagrams then turns into a geometric series, allowing us to perform the summation and write the answer in a compact form. 
 The ``renormalized'' kinetic equation  is the the sum of two terms, with $\mL_a$ and $\mL_b$:
\bea 
&\frac{\d n_k}{\d t} = -16\Im\sum\nolimits_{a,b} \sum\nolimits_{p_2,p_3, p_4}\!\! \lam_{k234}^2
\Big\{\hat\o[34; k2]\Big[ \frac{(n_k+n_2)n_3n_4}{1-\mL}\nonumber\\&- \frac{(n_3+n_4)n_kn_2}{1-\mL^*} \Big] 
 +  \mathcal{M}\frac{(n_k+n_2)(n_3+n_4)}{|1-\mL|^2}\Big\}=\tilde I_k~, \label{423}\\
& \mathcal{M}=2\sum\nolimits_{p_5} \frac{\lam_{k256}\lam_{5634}}{\lam_{k234}}{n_5 n_6  \hat\o[12; 56]\hat\o[34; 56]}\,.\nonumber\eea
where $\hat\o[ij;mn]\equiv \lim_{\eps\to0} (\omega_{ij;mn}+i\eps)^{-1}$. The equation 
 allows one to obtain several fundamental conclusions and suggests far-reaching assumptions about  weak and strong wave turbulence.

The renormalized kinetic equation is UV-convergent, that is valid at arbitrary $k_m$.  It is most likely that some corrections of higher orders will have UV-divergences cut off by the denominators $|1-\mL_a|^{-2}$ and $|1-\mL_b|^{-1}$. The cut-off then depends on $\lambda$, which will generally make the corrections proportional to non-integer power of $\lambda$, signaling non analiticity and singular renormalized perturbation theory, like for transport states. In any case, the cascade solution is independent of $k_m$ when $k_m\to\infty$.

This is not the case for the limit $k_0\to0$ even though the integrals over $p_2,p_3,p_4$ converge in the IR as well. The most important consequence of \eqref{423} is that the true dimensionless coupling is not $\epsilon_k$ from \eqref{39} but the loop integrals (\ref{aterm},\ref{bterm}) estimated as 
\be  \mL_a\simeq \mL_b\simeq \epsilon_0(k/k_0)^{\beta/2-\alpha}= \epsilon_k(k/k_0)^{\beta/6}\gg\epsilon_k\,,\label{var}\ee
Corrections to the weak-turbulence solution are proportional to $\mL$, which grows faster (or decay slower) along the direct cascade. When we keep the small parameter $\epsilon_k$ fixed and decrease $k_0$, we eventually violate the validity of the Zakharov solution and the kinetic equation \eqref{32}, which brings us to strong turbulence.

There are two alternative scenarios for the character of strong turbulence which are often discussed in the literature (a criterion for choosing between them for a given system isn't given). The first is a qualitatively similar cascade,  just with  the weak-turbulence time, $1/\omega_k\epsilon_k^2$ replaced by the nonlinear time $1/\omega_k\epsilon_k=1/\lambda_kn_kk^d$, so that the spectral energy flux $P$ is estimated as the spectral density $\tilde \omega_kn_kk^d$ divided by the nonlinear time,  $P\simeq \tilde \omega_kn_kk^d\lambda_kn_kk^d$, which gives  $n_k\simeq (P/\lambda_k\tilde\omega_k)^{1/2}k^{-d}$. We shall show below that the true answer is \eqref{strong1} instead. The second scenario is the often suggested hypothesis of critical balance $\epsilon_k=$const, which gives the universal (flux-independent) spectrum $n_k\simeq k^{-d}\omega_k/\lambda_k$, see e.g. \cite{Phil,GS,NZ,NS}. From the below consideration, we will see that the critical balance is $\mL(k)=$const instead. 

Let us now apply the renormalized kinetic equation \eqref{423} to the classes of turbulence with $\beta\geq 2\alpha$, where the effective coupling grows along the energy cascade. Now, instead of \eqref{gp}, the flux constancy requires
\be k^{d}\tilde\omega_k\tilde I_k={\rm const}\,.\label{flux1}\ee
The factor $|1-\mL|^{-1}$ determines the deviations from weak turbulence and the character of strong turbulence. Like in the field theory, the  sign of $\mL$ is crucial. 

A negative sign means that the multi-mode correlations suppress interactions like screening in quantum electrodynamics. Increase of $|\mL|$ (with increasing $k$ or decreasing $k_0$) must lead to an increase of $n_k$ relative to Zakharov spectrum to keep the same flux \eqref{flux1}. Upon further increase of  $|\mL|$ beyond unity, we put $ k^{d}\tilde\omega_k\tilde I_k\simeq k^{2\beta+3d}n_k^3/|\mL|=$const, which predicts the strong-turbulence spectrum falling with $k$ slower than \eqref{gp}:
\be n_k={k^{-d-2\beta/3}\over\mL^{1/3}}\simeq k^{-d-{2\beta\over3}}\left({k\over k_0}\right)^{\beta-2\alpha\over6}\epsilon_0^{-1/3}\,.\label{strong1}\ee

A positive sign of the effective coupling corresponds to interaction enhancement like in quantum chromodynamics, where the effective coupling  grows  with the nonlinearity parameter. Similarly, we expect no smooth passing through $\mL\simeq1$, but a (confinement) transition to strong turbulence dominated by bound states. Formally, the equation \eqref{flux1} could have a solution approaching at large $k$ the critical balance $|\mL|=1$.
Finding such solutions for specific systems will be attempted elsewhere. 

The above picture with a single sign of the effective coupling is the simplest case. Generally, the vertex renormalization depends on all four wavevectiors and may have different signs for different configurations. 
For example, the sign of the UV contribution to the one-loop vertex renormalization \eqref{UUVV} is determined 
by the product of the bare vertex and the second derivative of the  frequency (group velocity dispersion).
This is in stark contrast with  thermal equilibrium where different regimes are related to the sign of $\lambda$ --- the Gibbs state is non-normalizable when $\lambda<0$, which corresponds to attraction between waves. 
Our derivation shows that the bound states in turbulence could dominate not when there is attraction but when the signs of nonlinearity and dispersion are opposite, which is also the condition for solitons and collapses.
In the combination $k_ik_j\partial^2\omega_k/\partial k_i\partial k_j=k_\perp^2\omega'/k+k_\parallel^2\omega''$ we may have positive and negative terms. Strong turbulence will likely be determined only by positive contributions. When it is due to the interaction of parallel waves, this   will likely result in strong turbulence being dominated by quasi-one-dimensional objects like shocks or cusps. When the interaction of perpendicular waves is enhanced, one may expect turbulence dominated by solitons or collapse events.

{\bf Discussion}. Here we described how  nonlocality enhances or suppresses nonlinearity in non-equilibrium: deviations from weak-turbulence are magnified by IR divergences. The ratio of IR to UV scales is   essentially the non-equilibrium degree, analogous to the Reynolds number in hydrodynamics. How large Reynolds number enhances small nonlinearity was also   recently studied for shell models of turbulence \cite{FKV}.

The main technical result is the renormalized kinetic equation \eqref{423} where effective couplings are  (\ref{aterm},\ref{bterm}). Their signs determine whether one has enhancement or suppression of four-wave scattering by multi-wave correlations and the character of the transition from weak to strong turbulence. It is likely that universal (flux-independent) spectra dominated by bound states are possible when there is an enhancement. The hypothesis of structure-dominated universal spectra has, up to the present work, remained generic speculation, with neither a proof nor any criteria specifying when such spectra are possible. The interaction renormalization suggests such a criterium, which is an important step towards identifying universality classes of turbulence. Even more fascinating is the possibility that the new kinetic equation (\ref{423},\ref{flux1}) could describe both weak and strong turbulence (for instance, Zakharov and Phillips spectra of sea turbulence) as two opposite asymptotics of the steady casacade solution.

It is instructive to compare turbulence with the thermal-equilibrium equipartition spectra, $n_k=T/(\tilde\omega_k+\mu)$, which are independent of the form of the (weak) interaction. Weakly turbulent spectra depend only on the scaling exponents $\beta,\beta_1,\beta_2,\beta_3$ of the bare four-wave coupling $\lambda_{1234}$ (as well as on the dispersion relation and the spatial dimension). Since this work ventures outside of the domain of weak turbulence, there is more sensitivity to the specific form of  $\lambda_{1234}$.

Let us briefly discuss similarities and  differences between the relative roles of the largest and smallest scales known for fluid turbulence, and those found here for wave turbulence. In any turbulent direct cascade,  scale invariance is broken both by the pumping scale $L$ in the IR and the dissipation scale in the UV. In incompressible fluid turbulence, all the velocity moments are finite in the limit of the dissipation (viscous) scale going to zero. We identified here wide classes of wave turbulence with  the spectrum independent of a distant dissipation scale (however, the example of spin waves in the Appendix shows that wave turbulence allows for richer opportunities). Turning to large scales, the anomalous scaling in fluid turbulence makes the  effect of the pumping scale felt at  arbitrarily short scales  \cite{SF,FGV}: When one fixes the energy flux (the third velocity moment in incompressible fluid turbulence) at a given $k$, the spectral energy  (the second moment) at that scale goes to zero when $kL\to \infty$, while the moments higher than the third diverge. The larger the box, the more intermittent the system is, so that the same flux is transferred by more rare fluctuations.  Similarly, our IR divergence makes the corrections to weak wave turbulence explicitly dependent on the pumping scale. Even when the dimensionless nonlinearity parameter $\epsilon_k$ is small, by increasing $L=1/k_0$ one eventually breaks the weak-turbulence approximation and makes turbulence strong. The dependence of wave turbulence on the pumping scale discovered here is thus a direct analog of intermittency and  anomalous scaling.

Note that the space dimensionality $d$ influences neither the naive dimensionless parameter $\epsilon_k=\epsilon_0(k/k_0)^{\beta/3-\alpha}$, nor the effective one (in distinction, say, from passive-scalar turbulence \cite{FGV}). This is because the effective nonlinearity is determined by the total number of waves, $n_kk^d$, which does not depend on $d$ in turbulent cascades.
Recall that rigorous proofs of the validity of the bare kinetic equation are done in the double scaling limit $\lambda\to0$ and $L\to\infty$,
while keeping some combination finite  \cite{Math1,Math2,Math3}.  An important lesson from the present work is given by (\ref{V0}), which shows which combination of  $\lambda$ and $k_0=1/L$ defines corrections to the weak-turbulence spectra of a direct energy cascade.

We thank M. Smolkin for collaboration on related work, A. Zamolodchikov, J. Shatah, D. Schubring, and C. Cheung for helpful discussions.  The work of VR is supported by NSF grant 2209116 and by the ITS through a Simons grant. VR thanks the Aspen Center for Physics (NSF grant  PHY-2210452) and  GF thanks NYU and the Simons Center for Geometry and Physics where part of this work was completed. The work of GF is  supported by the Excellence Center at WIS, by the grants  662962 and 617006 of the Simons  Foundation,    and by the EU Horizon 2020 programme under the Marie Sklodowska-Curie grant agreements No 873028 and 823937.

\newpage
\onecolumngrid
\appendix

\section{Derivation of higher order loops diagrams}\label{sec:Renorm}
In this appendix we present the derivation of the higher order (loop diagram) contributions to the four-point function,  which determines the ``renormalized'' kinetic equation  through the relation (\ref{424v2}).
We will obtain the four-point function by modifying our equations of motion to add dissipation and Gaussian-random forcing to the system:
\be \frac{da_k}{dt}=-i\partial H/\partial a_k^*+f_k(t)-\gamma_ka_k\,.\label{e1}\ee The force is taken to be white in time and acting on each mode independently: $\langle f_k(0) f_q(t)\rangle =F_k\delta(k{-}q)\delta(t)$. At the end of the computation, one takes the limit $F_k,\gamma_k\to0$, while keeping $F_k/2\gamma_k=n_k$ finite. The respective standard Wyld diagram technique was introduced in \cite{Wyld1}, applied to wave turbulence in \cite{Wyld2}, and given a modern compact form in \cite{RS1}. Acting by an independent random force $f_k$ on every mode is natural in thermal equilibrium, where one uses the equipartition occupation numbers, $n_k\omega_k=$const. Starting from \cite{Wyld2}, one also uses this way of averaging for arbitrary $n_k$. There are always factors in the environment that scramble phase correlations; modeling them by an infinitesimal white force is an artificial yet legitimate option. It is reinforced by the fact that the kinetic equation obtained this way does have turbulent  Zakharov exact solutions, described below.
\begin{figure} \centering
\includegraphics[width=1.2in]{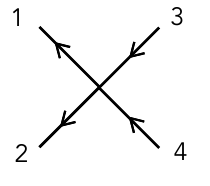}
\caption{The leading order kinetic equation (\ref{32}) encodes the tree-level process of two modes directly scattering into two other modes. } \label{Ptree}
\end{figure}

In such a straightforward (naive) perturbation theory, one takes the temporal Fourier transform and writes the solution of (\ref{e1}) as a series in $\lambda$. Denoting $q=(k,\omega)$ one gets:
\bea && [i(\omega_k{-}\omega)+\gamma_k] a_q=f_q-i\int {dq_1dq_2dq_3\over(2\pi)^2}\,\delta(q{+}q_1{-}q_2{-}q_3)\lambda_{k123}a_{q_1}^*a_{q_2}a_{q_3}\nonumber\\
&\approx& f_q-i\int {dq_1dq_2dq_3\over(2\pi)^2}\delta(q{+}q_1{-}q_2{-}q_3){\lambda_{k123}f_{q_1}^*\over \gamma_1-i(\omega_{k_1}\!-\omega_1)}\,  {f_{q_2}\over i(\omega_{k_2}\!-\omega_2)+\gamma_2}\, {f_{q_3}\over i(\omega_{k_3}\!-\omega_3)+\gamma_3}+O(\lambda^2)\,. \nonumber
\eea
Multiplying such expressions and averaging over $f_q$, one obtains the first non-vanishing contribution to the fourth moment in (\ref{424v2}), which could be symbolically depicted as the tree-level diagram shown in Figure~\ref{Ptree}. This
 gives the standard kinetic equation in the main text. 

A more effective method for doing such calculations was introduced in \cite{RS1}, and this is what we will use. In particular, \cite{RS1} showed that our classical stochastic field theory  is equivalent to a quantum field theory, with  expectation values given by a path integral for a Lagrangian that is the square of the classical, force-free, equations of motion ${E}_{f=0} = \dot a_k + i \frac{\delta H}{\delta a^*_k} +\gamma_k a_k$:
\be \label{Leff}
\la \mO\ra = \int  \mathcal D a \mathcal D a^* \mO(a)   e^{-\int dt L}\,, \ \  L = \sum_k \frac{|E_{f=0}|^2 }{F_k} ~.
\ee
Here $F_k$ is the variance of the Gaussian-random forcing and $\gamma_k$ is the dissipation.
The limit that we will take, after computing the correlation functions, is $F_k, \g_k \rightarrow 0$ with fixed $n_k$:
$n_k = F_k/2 \g_k$.
The reason for this notation is that $n_k$ is the occupation number of mode $k$ for the noninteracting theory.

\subsection*{Feynman rules}
The Lagrangian contains quadratic, quartic, and sextic terms. As usual, the Feynman rules are most convenient in  momentum-frequency space.
The quadratic term in the Lagrangian  gives the propagator,
\be \label{DG}
D_{k, \o}  =2\g_k n_k |G_{k, \o}|^2~, \ \ \ \ \   \text{where} \ \ \ \ \ \ G_{k, \o} = \frac{i}{\o{-} \o_{k} + i \gamma_{k}}~.
\ee
We will use shorthand $D_i \equiv D_{p_i, \o_i}$, and $\lam_{p_1 p_2 p_3 p_4} \equiv \lam_{1234}$,
and it is also convenient to define,
\be
g_i = \frac{1}{G_{p_i, \o_i} 2\g_{p_i} n_{p_i}}~.
\ee
Notice that in the limit of $\g_k\rightarrow 0$, the propagator becomes,
\be \label{56}
D_{k,\o}  =  n_k \frac{2\g_k}{(\o - \o_k)^2 + \g_k^2} \rightarrow  n_k 2\pi \delta(\o- \o_k)~.
\ee
It will, however, be important to keep $\g_k$ finite until the very end of the calculation. The Feynman rule for the quartic interaction comes with a factor
\be \label{quarticV2}
 - i \lam_{1234} (g_1^* {+} g_2^* {-} g_3 {-} g_4) \, 2\pi \delta(\o_{12;34})~,
\ee
where we have introduced the notation $\o_{12;34} \equiv \o_1{+}\o_2{-}\o_3{-}\o_4$.

The  equal-time four-point function  in (\ref{424v2}) is given by a Fourier transform of the on in the frequency domain:
\be
\la a_{p_1}(t) a_{p_2}(t) a^*_{p_3}(t) a^*_{p_4}(t) \ra = \int \frac{d\o_1}{2\pi} \cdots\frac{ d \o_4}{2\pi}  e^{i (\o_3{+}\o_4{-}\o_1{-}\o_2)t }\la a_{p_1, \o_1} a_{p_2, \o_2} a^*_{p_3, \o_3} a^*_{p_4, \o_4} \ra~.\label{FT4}
\ee
The tree-level contribution to the frequency-space four-point function, shown earlier in Fig.~\ref{Ptree},  follows immediately from the Feynman rule for the quartic vertex (\ref{quarticV2}). We simply attach external propagators and add a combinatorial factor of four, to get,
\be\label{4ptF}
L_0(1,2,3,4) = - 4 i \lam_{1234}
(g_1^* {+} g_2^* {-} g_3 {-} g_4) \prod_{i=1}^4 D_i \, 2\pi \delta(\o_{12;34})~,
\ee
where we will be using $L(1,2,3,4)$ with various subscripts to denote different contributions to the frequency-space four-point function.

The one-loop Feynman diagram consists of a single bubble \cite{RS1}:
\be
 L_1(1,2,3,4)={-}16\pi \delta(\o_{12;34})\prod_{i=1}^4 D_i \sum_{p_5} \int \frac{d\o_5}{2\pi} D_5 D_6  \lam_{1256}\lam_{5634}(g_1^* {+} g_2^* {-} g_5{-}g_6)(g_5^* {+} g_6^*{-} g_3{-}g_4)\,, \label{55}
\ee
where $\o_6 = \o_3 {+}\o_4{-} \o_5$ and $p_6 = p_3{+}p_4{-}p_5$. Note that the coupling $\lam_{1234}$ contains a momentum conserving delta function $\delta(p_1{+}p_2{-}p_3{-}p_4)$ which we keep implicit. To get this, we simply applied the Feynman rule for the vertex (\ref{quarticV2}) to each of the two vertices, wrote a propagator $D_i$ for each of the lines, and wrote a sum for the internal momentum ($p_5$) and frequency ($\o_5$). The procedure for opposite arrows is the same.

That gives the contribution to $\partial n_k/ \partial t$ given in the main text\cite{RS1, RS2}:
 \be
64\pi \sum_{p_1, \ldots, p_6}  \delta(\o_{k2;34} )\prod_{i=2}^6  n_i \, \Big(\frac{1}{n_k} {+} \frac{1}{n_2}{-}\frac{1}{n_3} {-} \frac{1}{n_4} \Big)\left[
    \frac{ \lam_{34k2} \lam_{k256}\lam_{5634}  }{\o_{k256}}\Big(\frac{1}{n_5} {+} \frac{1}{n_6} \Big)+4  \frac{\lam_{k234} \lam_{k645}\lam_{2536}  }{\o_{3625}} \Big(\frac{1}{n_5} {-} \frac{1}{n_6} \Big)\right]\,.\label{bterm}
\ee
The integral of $1/\o$ is the principal value part.

\begin{figure}[t]
\centering
\subfloat[]{\includegraphics[width=1.7in]{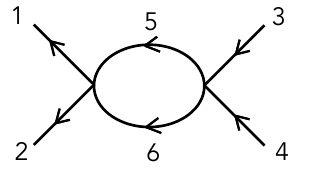}} \ \ \  \ \  \ \  \ \ \ \  \ \ \
\subfloat[]{\includegraphics[width=1.7in]{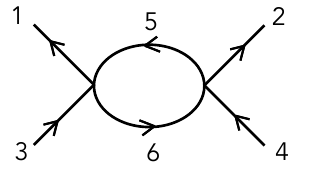}}
\caption{The $\lambda^3$-terms in the kinetic equation, (\ref{aterm}) and (\ref{bterm}). 
 } \label{Ploop}
\end{figure}

\subsection{Two loop contribution to the fourth moment}
\begin{figure}[h]
\centering
\includegraphics[width=2in]{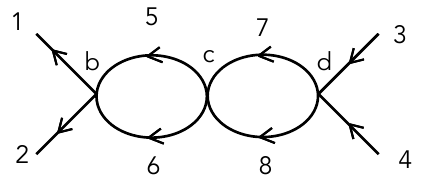}
\caption{ }
\label{P4loop}
\end{figure}
The one-loop correction  to the equal-time four-point function is given in the main text. Let us analyze here the two-loop  diagrams looking for the leading asymptotics in the UV. We start from the two-loop bubble diagrams shown in Fig.~\ref{P4loop}. The contribution of the term with all arrows in the same direction is as follows:
\be\label{B39}
-16\sum_{p_5,p_7} \frac{\lam_{1256} \lam_{5678}\lam_{7834}}{\o_{p_3p_4;p_1p_2}{+}i\eps} \prod_{i}^{8}n_i\(T_{bcd}+T_{cdb} + T_{cbd}+T_{dcba}\)~,
\ee
where
\bea \nn
T_{bcd} &=&\frac{i}{\o_{p_5p_6;p_1p_2}{+}i\eps}\frac{i}{\o_{p_7p_8;p_1p_2}{+}i\eps} \Big(\frac{1}{n_5}{+}\frac{1}{n_6}{-}\frac{1}{n_1}{-}\frac{1}{n_2}\Big)\Big(\frac{1}{n_7}{+}\frac{1}{n_8}\Big) \Big(\frac{1}{n_3}{+}\frac{1}{n_4}\Big)   \\ \nn
T_{cdb} &=& \frac{i}{\o_{p_7p_8;p_5p_6}{+}i\eps} \frac{i}{\o_{p_3p_4;p_5p_6}{+}i\eps}\Big(\frac{1}{n_7}{+}\frac{1}{n_8}{-}\frac{1}{n_5}{-}\frac{1}{n_6}\Big)\Big(\frac{1}{n_3}{+}\frac{1}{n_4}\Big) \Big(\frac{-1}{n_1}{-}\frac{1}{n_2}\Big) \\ \nn
T_{cbd} &=& \frac{i}{\o_{p_7p_8;p_5p_6}{+}i\eps}  \frac{i}{\o_{p_7p_8;p_1p_2}{+}i\eps} \Big(\frac{1}{n_7}{+}\frac{1}{n_8}{-}\frac{1}{n_5}{-}\frac{1}{n_6}\Big)\Big(\frac{-1}{n_1}{-}\frac{1}{n_2}\Big) \Big(\frac{1}{n_3}{+}\frac{1}{n_4}\Big)    \\
T_{dcb} &=&  \frac{i}{\o_{p_3p_4;p_7 p_8}{+}i\eps}  \frac{i}{\o_{p_3p_4;p_5 p_6}{+}i\eps}  \Big(\frac{1}{n_3}{+}\frac{1}{n_4}{-}\frac{1}{n_7}{-}\frac{1}{n_8}\Big)\Big(\frac{-1}{n_5}{-}\frac{1}{n_6}\Big) \Big(\frac{-1}{n_1}{-}\frac{1}{n_2}\Big)~,\label{2a}
\eea
where the $T_{bcd}$ term comes form the time ordering $t_b{>}t_c{>}t_d{>}t_a$, $T_{cdb}$  comes form the time ordering $t_c{>}t_d{>}t_b{>}t_a$, and so on ($t_a$, which is the time at which the correlation function is evaluated, is always the earliest time). 

It produces the IR divergence $k_0^{4(\beta/4-\beta/3)}$ when $p_5,p_7\to 0$. In this case 
$T_{cdb}+T_{cbd}\to0$ and $T_{bcd}+T_{dcb}\propto n_1n_2n_3n_4\lambda_{12,1+2}^2\lambda |p_1+p_2|^{2\beta/2}k_0^{4(\beta/4-\beta/3)}(\omega_1+\omega_2-\omega_{1+2})^{-2}(1/n_3+1/n_4-1/n_1-1/n_2)$.

The respective contribution of the term with opposite directions of the arrows in the bubbles is obtained by replacing $2\leftrightarrow -4$,  $6\leftrightarrow -6$,  $8\leftrightarrow -8$:
\be\label{B39b}
-16 \sum_{p_5,p_7}\frac{\lam_{1645} \lam_{5867}\lam_{7283}}{\o_{p_3p_4;p_1p_2}{+}i\eps}\prod_{i}^{8}n_i\sum T'
\ee
where
\bea \nn
T'_{bcd} &=&\frac{i}{\o_{p_5p_4;p_1p_6}{+}i\eps}\frac{i}{\o_{p_7p_4;p_1p_8}{+}i\eps} \Big(\frac{1}{n_5}{+}\frac{1}{n_4}{-}\frac{1}{n_1}{-}\frac{1}{n_6}\Big)\Big(\frac{1}{n_7}{-}\frac{1}{n_8}\Big) \Big(\frac{1}{n_3}{-}\frac{1}{n_2}\Big)   \\ \nn
T'_{cdb} &=& \frac{i}{\o_{p_7p_6;p_5p_8}{+}i\eps} \frac{i}{\o_{p_3p_6;p_5p_2}{+}i\eps}\Big(\frac{1}{n_7}{+}\frac{1}{n_6}{-}\frac{1}{n_5}{-}\frac{1}{n_8}\Big)\Big(\frac{1}{n_3}{+}\frac{1}{n_4}\Big) \Big(\frac{-1}{n_1}{-}\frac{1}{n_2}\Big) \\ \nn
T'_{cbd} &=& \frac{i}{\o_{p_7p_6;p_5p_8}{+}i\eps}  \frac{i}{\o_{p_7p_2;p_1p_8}{+}i\eps} \Big(\frac{1}{n_7}{+}\frac{1}{n_6}{-}\frac{1}{n_5}{-}\frac{1}{n_8}\Big)\Big(\frac{-1}{n_1}{+}\frac{1}{n_4}\Big) \Big(\frac{1}{n_3}{-}\frac{1}{n_2}\Big)    \\
T'_{dcb} &=&  \frac{i}{\o_{p_3p_8;p_7 p_2}{+}i\eps}  \frac{i}{\o_{p_3p_6;p_5 p_2}{+}i\eps}  \Big(\frac{1}{n_3}{+}\frac{1}{n_8}{-}\frac{1}{n_7}{-}\frac{1}{n_2}\Big)\Big(\frac{-1}{n_5}{+}\frac{1}{n_6}\Big) \Big(\frac{-1}{n_1}{+}\frac{1}{n_4}\Big)~.\label{2b}
\eea
Again, we find similar cancellations, so that the main contributions come from the first and last terms in (\ref{2a},\ref{2b}): 
\bea &T_{bcd}+T_{dcb}+T'_{bcd}+T'_{dcb}\simeq {16\lambda^3 k_0^{2\beta/3-2\beta/2}\over \o_{p_3p_4;p_1p_2}{+}i\eps} \nonumber\\&\times
\left[{|p_2+p_1|^{2\beta/2}\over(\omega_2+\omega_1-\omega_{2+1})^{2}}
+{|p_2-p_3|^{2\beta/2}\over (\omega_2-\omega_3-\omega_{2-3})^{2}}\right]\nonumber\\&\times n_1n_2n_3n_4(1/n_3+1/n_4-1/n_1-1/n_2)
\,.\label{2UV}
\eea
The main question now is the divergence degree at UV relative to the bare kinetic equation (KE). We have two extra vertices which bring $k_m^{2\beta/2}$ and two extra combinations of the frequences, which can bring $-2\alpha$, $-\al-\kappa$ or $-2\kappa$.
We see that \eqref{2UV} contan no extra cancellations so that the power at UV is KE + $2(\beta/2-\kappa)$. 

\subsubsection{Other two-loop diagram}
Let us look at  the other  two-loop  diagram, \begin{figure}[h] \centering
\includegraphics[width=1.5in]{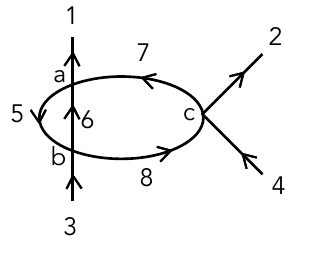}
\caption{}  \label{IRlam3bv3}
\end{figure}
 The result is a sum over the different time orderings:
\be
\sum_{5,\ldots,8}\frac{ \lam_{1567} \lam_{5386}\lam_{7284}}{\o_{p_3p_4;p_1p_2}{+}i\eps}\prod_{i=1}^8 n_i\sum F
\ee
where
\bea
F_{abc} &\rightarrow& \ho\[67;15\]\hn\[67;15\]\ho\[73;18\]\hn\[3;8\]\hn\[4;2\] \\
F_{acb}&\rightarrow&\ho\[67;15\]\hn\[67;15\]\ho\[468;215\]\hn\[48;2\]\hn\[3;\]\\
F_{bac}&\rightarrow&\ho\[35;68\]\hn\[35;68\]\ho\[37;18\]\hn\[7;1\]\hn\[4;2\]\\
F_{bca}&\rightarrow&\ho\[35;68\]\hn\[35;68\]\ho\[345;267\]\hn\[4;27\]\hn\[;1\]\\
F_{cab}&\rightarrow&\ho\[48;27\]\hn\[48;27\]\ho\[468;125\]\hn\[6;15\]\hn\[3;\]\\
F_{cba}&\rightarrow&\ho\[48;27\]\hn\[48;27\]\ho\[345;267\]\hn\[35;6\]\hn\[;1\]
\eea
Here we consider $p_5,p_6\to0$ and  $p_1,p_3,p_8,p_7\to \infty$. The  two diagrams are subleading:
$$F_{abc}+F_{bac}=\ho\[67;15\]\hn\[67;15\]\ho\[73;18\]\hn\[73;18\]\ho\[34;12\]\hn\[4;2\] \propto (\gamma/\alpha)^2p_m^{2\gamma-2\alpha-\kappa}\,.$$
\bea
&F_{acb}\rightarrow {\gamma\over\alpha}p_m^{\gamma-\alpha}\ho\[34;12\]\hn\[34;2\]\hn\[3;\]\rightarrow{\gamma\over\alpha}p_m^{2\gamma-\alpha}\ho\[34;12\]\hn\[34;2\]\\
&F_{bca}\rightarrow {\gamma\over\alpha}p_m^{\gamma-\alpha}\ho\[34;12\]\hn\[4;12\]\hn\[;1\]\rightarrow-{\gamma\over\alpha}p_m^{2\gamma-\alpha}\ho\[34;12\]\hn\[4;12\]\\
&F_{cab}\rightarrow \ho^2\[34;12\]\hn\[34;12\]\hn\[;1\]\hn\[3;\]\rightarrow p_m^{2\gamma}\ho^2\[34;12\]\hn\[34;12\]\\
&F_{cba}\rightarrow \ho^2\[34;12\]\hn\[34;12\] \hn\[3;\] \hn\[;1\]\rightarrow p_m^{2\gamma}\ho^2\[34;12\]\hn\[34;12\]\\
\eea
\bea &\lim_{p_2,p_4\to\infty}\Im\ho\[34;12\]\hn\[34;12\]=\lim_{p_2,p_4\to\infty}\delta(\o_{12;34})\hn\[34;12\]\simeq {\gamma \over \alpha} p_2^{\gamma-\alpha-\kappa}\label{lim0}\\
&\lim_{p_2,p_4\to\infty}\Im\ho^2\[34;12\]\hn\[34;12\]=-\delta(\o_{12;34})\partial_\omega\hn\[34;12\]=-{\gamma(\gamma-\alpha)\over\alpha^2}p_2^{\gamma-2\alpha-\kappa}\,.\label{lim1}\\
&\lim_{p_2,p_4\to\infty}\Im\ho^3\[34;12\]\hn\[34;12\]=-{1\over 2}\delta(\o_{12;34})\partial_\omega^2\hn\[34;12\]=-{\gamma(\gamma-\alpha)(\gamma-2\alpha)\over2\alpha^3}p_2^{\gamma-3\alpha-\kappa}\,,\label{lim2}\\
&\lim_{p_2/p_1\to\infty}\ho\[1+2;12\]\hn\[1+2;12\]={\gamma\over\alpha}p_2^{\gamma-\alpha}\,, \ \alpha\geq1\,.\label{lim3}\eea
and obtain
 \bea&\sum F=2p_m^{2\gamma}\ho^2\[34;12\]\hn\[34;12\]+{\gamma\over\alpha}p_m^{2\gamma-\alpha}\ho \[34;12\]\hn\[3;1\]\nonumber\\
&=- {\gamma(\gamma-\alpha)(\gamma-2\alpha)\over \alpha^3}p_2^{3\gamma-3\alpha}-{\gamma^2(\gamma-\alpha)\over\alpha^3}p_2^{3\gamma-3\alpha}\,.\nonumber\eea
After multiplying by $\lambda^3k_0^{4(\beta/4-\beta/3)}p_m^{\beta-\beta/2+3\beta/2}\delta(o_{34;12} )\prod_{i=1}^4 n_i$, the sum of the four diagrams gives $\propto 2(\gamma/\alpha)p_m^{2\beta/2-2\beta/3-\alpha-\kappa+2\beta/2-2\alpha}$, that is the power KE+$2\beta/2-2\alpha$, which is subleading relative to $\sum T$ which gave   KE+$2\beta/2-2\kappa$. Indeed, $\alpha>\kappa=\max\{0,\al-1\}$.

We need to add the same diagram with a different choice of arrows, as shown in  in Fig.~\ref{IRlam3bv2}, which is Fig.~\ref{IRlam3bv3}   but with $ 2\leftrightarrow -3$ and $8\leftrightarrow -8$.
\begin{figure}[h] \centering
\includegraphics[width=1.5in]{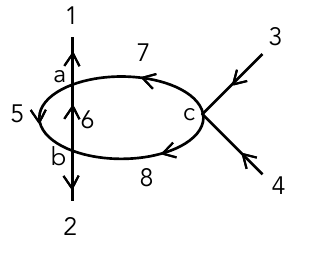}
\caption{}  \label{IRlam3bv2}
\end{figure}

\be
\sum_{5,\ldots,8}\frac{\lam_{1567} \lam_{5826}\lam_{7834}}{\o_{p_3p_4;p_1p_2}{+}i\eps} 
\prod_{i=1}^8 n_i\sum G
\ee
where
\bea
G_{abc} &=&\ho\[67;15\]\hn\[67;15\]\ho\[78;12\]\hn\[8;2\]\hn\[34\]\nonumber\\
G_{bac}&=&\ho\[58;26\]\hn\[58;26\]\ho\[78;12\]\hn\[7;1\]\hn\[34\]\nonumber\\
G_{acb}&=&\ho\[67;15\]\hn\[67;15\]\ho\[26;58\]\hn\[34;8\]\hn\[;2\]\nonumber\\
G_{cab}&=&\ho\[34;78\]\hn\[34;78\]\ho\[26;58\]\hn\[6;15\]\hn\[;2\]\nonumber\\
G_{bca}&=&\ho\[58;26\]\hn\[58;26\]\ho\[15;67\]\hn\[34;7\]\hn\[;1\]\nonumber\\
G_{cba}&=&\ho\[34;78\]\hn\[34;78\]\ho\[15;67\]\hn\[5;26\]\hn\[;1\]\nonumber
\eea
This diagram gives the IR divergence $k_0^{4(\beta/4-\beta/3)}$ when $p_6,p_8\to 0$. Consider in addition $p_2,p_4,p_5,p_7\to \infty$, then $G_{bca},G_{cba}$ are negligible. What remains
\bea
G_{abc} &=&\ho\[1+2;12\]\hn\[1+2;12\]\ho\[1+2;12\]\hn\[;2\] \hn\[34\]\nonumber\\
G_{bac}&=&{\gamma\over\alpha}p_m^{\gamma-\alpha}\ho\[1+2;12\]\hn\[1+2;1\] \hn\[34\]\nonumber\\
G_{acb}&=&\ho\[1+2;12\]\hn\[1+2;12\]\ho\[34;12\]\hn\[34;\]\hn\[;2\]\nonumber\\
G_{cab}&=&\ho\[34;3+4\]\hn\[34;3+4\]\ho\[34;12\]\hn\[;12\]\hn\[;2\] \nonumber
\eea
\bea &G_{acb}+G_{cab} = -p_m^{2\gamma}\ho\[1+2;12\]\ho\[34;12\]\hn\[34;12\]\nonumber\\&\sim-{\gamma(\gamma-\alpha)\over\alpha^2}p_2^{3\gamma-2\alpha-\kappa} \,.\label{Gsub}\eea
The contribution (\ref{Gsub}) is subleading comparing to the following one (written for $\alpha>1$):
\bea 
&G_{abc}+G_{bac}=\ho\[12,1+2\]\hn\[12,1+2\] {\gamma\over\alpha}p_m^{2\gamma-\alpha}\nonumber\\
&\simeq (\gamma/\alpha)^2p_m^{3\gamma-2\alpha}\,.\label{problem}\eea
After multiplying by $\lambda^3 n_1n_2 n_3n_4  k_0^{2(\beta/3-\beta/2)}p_m^{2\beta/2}\delta(\o_{1234})$ the power of $p_m$ is KE +$2\beta/2-\alpha$. It is subleading for $\alpha<2$ comparing to  $T$ that gives KE +$2\beta/2-2(\alpha-1)$. So we see no extra cancellations in UV asymptotics of two-loop contributions into the kinetic equation.

We just come to the conclusion that the leading UV divergences correspond to two-loop bubble diagrams $T$ and $T'$. Note that there is no cancellations at UV for two-loop bubble diagrams, similar to cancellations between one-loop diagrams. Assuming that the same situation takes place at higher orders, we come to the conclusion that the main sequence of UV divergences choose two geometric series of bubble diagrams, of a and b types.


\subsection{Bubble diagrams}

As mentioned, to cure the UV divergences at one loop, we need to go beyond one loop and sum diagrams with an arbitrary number of bubbles, as shown in Fig.~\ref{bubblesum}. These constitute a subset of the most UV divergent diagrams at each order. We will discuss the bubble diagrams in which the arrows all go in the same direction. The modification for the diagrams in which the arrows go in the opposite direction is straightforward. In fact, the sum of bubble diagrams with arrows in opposite direction constitutes the full answer (at large $N$) for a theory with $N$ fields, as opposed to one field, like we have been discussing, see \cite{Berges, RS}. 
\begin{figure}[h]
\centering
\includegraphics[width=4in]{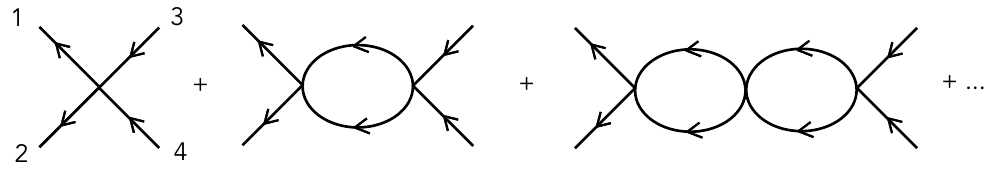}
\caption{ }  \label{bubblesum}
\end{figure}

As a digression,  we note that the same sequence of bubble diagrams  appears in many other contexts. It appears in the calculation of the correlation energy of an electron gas, as the most IR divergent diagrams. In this context these diagrams are referred to as sausage diagrams \cite{Feynman, GellMann}. The same diagrams appear in the computation of Kawasaki and Oppenheim \cite{KO} for  transport coefficients, as the most IR divergent diagrams, where they are referred to as ring diagrams. Their sum in this context gives a log term, and constitutes the derivation of the Dorfman-Cohen effect. Note that in these two (particle) systems each individual diagram is an IR divergent space  integral, whereas in the context of waves (this paper) the Fourier-space integrals are UV divergent. The main difference lies in the quartic interaction $\lam_{1234}$ appearing in the loop integral. In the context of waves, $\lam_{1234}$ grows with momentum, i.e. $\beta/2$ is positive, and when $\beta/2> \kappa$ we have the UV divergence discussed in this paper. For particles, on the other hand, the interaction either decays at large momenta or is independent of the magnitude of momentum, as for hard spheres.
\begin{figure}[h]
\centering
\includegraphics[width=2.4in]{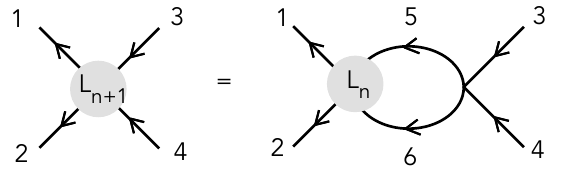}
\caption{ }  \label{SD2}
\end{figure}

Proceeding, we denote the diagram with $n$ bubbles, with all arrows running in the same direction, by $L_n(1,2,3,4)$ and compute the sum,
\be
L(1,2,3,4) = \sum_{n=0}^{\infty} L_n(1,2,3,4)~.\label{sum}
\ee
We can obtain the diagram with $n{+}1$ bubbles by attaching a loop to the the diagram with $n$ bubbles, as shown below in Fig.~\ref{SD2}.
\begin{figure}[h]
\centering
\includegraphics[width=2.4in]{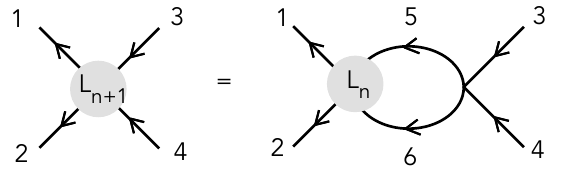}
\caption{ }  \label{SD2}
\end{figure}
Explicitly, we have that,
\be \label{513}
L_{n+1}(1,2,3,4) =  - 2i D_3 D_4 2\pi\delta(\o_{12;34})\sum_{p_5,p_6} \int \frac{d\o_5}{2\pi} \frac{d\o_6}{2\pi}  L_{n}(1,2,5,6) \lam_{5634} (g_5^* {+} g_6^*{-} g_3{-}g_4)~.
\ee
 Starting with $L_0$ given earlier in (\ref{4ptF}), we may iteratively compute $L_n$.

The sum $L(1,2,3,4)$ of all the bubble diagrams satisfies the integral (Schwinger-Dyson) equation
\be \label{SDeqn}
L(1,2,3,4) = L_0(1,2,3,4)  - 2i D_3 D_4 2\pi\delta(\o_{12;34})\sum_{p_5,p_6} \int \frac{d\o_5}{2\pi}\frac{d\o_6}{2\pi}  L(1,2,5,6) \lam_{5634} (g_5^* {+} g_6^*{-} g_3{-}g_4)~,
\ee
as one can see by iterating this equation, represented pictorially in Fig.~\ref{SD}.
\begin{figure}[h]
\centering
\includegraphics[width=4in]{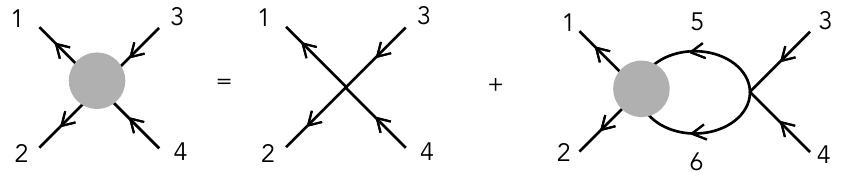}
\caption{ }  \label{SD}
\end{figure}

\subsection{Product-factorized couplings}
For general couplings $\lam_{1234}$ the integrals over the internal momenta of the loops do not factorize, making it difficult for us to write the solution of (\ref{SDeqn}) in a useful form.
Here we will consider the special case in which the ratio of couplings
\be \label{lamR}
\frac{\lam_{1256}\lam_{5634}}{\lam_{1234}}
\ee
is only a function of $p_5$ and $p_6$. To achieve this we take the couplings to have product factorization, $\lam_{1234} = \sqrt{\lam_{12}\lam_{34}}$ where $\lam_{12}$ depends on only $p_1$ and $p_2$, and so (\ref{lamR}) is $\lam_{56}$.
The sum of bubble diagrams then turns into a geometric series, allowing us to perform the summation and write the answer in a compact form. Indeed, one can, verify that  (\ref{SDeqn}) has the  solution,
\be \label{RKE}
L(1,2,3,4) =- 4 i \lam_{1234}
 \prod_{i=1}^4 D_i \, 2\pi \delta(\o_{12;34})
\Big\{
 \frac{(g_1^* {+} g_2^*)}{1 - N_1}
-\frac{( g_3 {+} g_4)}{1 - N_1^*}
+  \frac{(g_1^* {+} g_2^*) ( g_3 {+} g_4)}{1- N_1}\frac{M_1}{1- N_1^*}\Big\}~,
\ee
where we defined,
\be\label{NM2}
N_1 = \sum_{p_5}\lam_{56}  \frac{2 (n_5 {+} n_6)}{ \o_{12;p_5p_6} {+} i \gamma_{56}}~,\ \ \ \ \ \ \ \
M_1 =2i\sum_{p_5} \lam_{56} \frac{ 2n_5 n_6 \g_{56}}{\o_{12;p_5p_6}^2 {+} \gamma_{56}^2}~,
\ee
and introduced the notation $\g_{56} \equiv \g_5{+}\g_6$ (we will sometimes simply replace various sums of $\g_i$ by $\eps$, since it is equivalent). Momentum conservation fixes  $p_6 = p_1{+}p_2{-}p_5$, and one should make this replacement for $p_6$ in all places that it appears.

In slightly more detail, in evaluating the contour integral on the right-hand side of (\ref{513}), an integral of the following form   appears (see (D.1) of \cite{RS1}),
\bml \label{OneloopInt0}
\int \frac{d\o_5}{2 \pi} \(A+ B(g_5 {+} g_6)\) \(g_5^* {+} g_6^* {-} g_3 {-} g_4\) D_5 D_6\\
=  A  N(5,6) +\(A M(5,6)- B N(5,6)^* \)(g_3 {+}g_4)
+B \(\frac{n_6}{2 \g_5 n_5} {+}\frac{n_5}{2\g_6 n_6}\)~,
\end{multline}
for some constants $A, B$, and
where $\o_6 = \o_3 {+}\o_4{-} \o_5$ and we defined,
\be \label{NM}
N(5,6) = \frac{i (n_5 {+} n_6)}{ \o_{34;p_5p_6} {+} i \gamma_{56}}~, \ \ \ \ \ \ \
M(5,6) = -\frac{ 2n_5 n_6 \g_{56}}{\o_{34;p_5p_6}^2 {+} \gamma_{56}^2}~.
\ee
Since we will always have a frequency conserving delta function, $\delta(\o_{12;34})$, we can replace $\o_3{+}\o_4$ with $\o_1{+}\o_2$,
\be
N(5,6) = \frac{i (n_5 {+} n_6)}{ \o_{12;p_5p_6} {+} i \gamma_{56}}~, \ \ \ \ \ \ \
M(5,6) = -\frac{ 2n_5 n_6 \g_{56}}{\o_{12;p_5p_6}^2 {+} \gamma_{56}^2}~.
\ee
The last term in (\ref{OneloopInt0}) is divergent in the limit of vanishing dissipation, however it will always get canceled by diagrams involving the sextic interaction. In particular, the sextic term in the Lagrangian simply serves to cancel the contribution coming from the collision (in time) of the two vertices in the one-loop diagram \cite{RSSS}.
So, in effect, we can simply drop this term and write,
\be \label{OneloopInt}
\int \frac{d\o_5}{2 \pi} \(A + B(g_5 {+} g_6)\) \(g_5^* {+} g_6^* {-} g_3 {-} g_4\) D_5 D_6
``=" A  N(5,6) +\(A M(5,6)- B N(5,6)^* \)(g_3 {+}g_4)~.
\ee
Using (\ref{OneloopInt}) one can now easily check that (\ref{RKE}) satisfies (\ref{SDeqn}).
\\

We  now Fourier transform  (\ref{RKE}), and insert the resulting equal-time four-point function into (\ref{424v2}) to obtain  the ``renormalized'' kinetic equation, 
\bml 
\frac{d n_1}{d t} = 4\sum_{2,3,4} \lam_{1234}^2\prod_{i=1}^4 n_i\, \text{Im}\Big(  \frac{1}{\o_{p_3p_4; p_1p_2} {+} i\eps}\Big[ \Big(\frac{1}{n_1} {+} \frac{1}{n_2}\Big) \frac{1}{1-\mL_1(p_3, p_4) }- \Big(\frac{1}{n_3} {+} \frac{1}{n_4}\Big)\frac{1}{1-\mL_1(p_1, p_2)^*} \Big]\\
+\Big(\frac{1}{n_1}{+}\frac{1}{n_2}\Big)\Big(\frac{1}{n_3}{+}\frac{1}{n_4}\Big)\,  \sum_{p_5} \lam_{56}  \frac{2n_5 n_6 }{(\o_{p_5p_6; p_1 p_2} {+} i\eps)(\o_{p_5 p_6; p_3p_4} {-} i\eps)}  \frac{1}{|1-\mL_1(p_5, p_6) |^2}\Big)
 \end{multline}
 where $\mL_1 $ is 
  \be
\mL_1(p_3, p_4)= \sum_{p_5}\lam_{56}  \frac{2 (n_5 {+} n_6)}{\o_{p_3p_4;p_5p_6} {+} i \eps}~.\label{defL}
 \ee
In fact, one can simplify this expression to get \cite{RS},
\be
\frac{\d n_1}{\d t} = - 4\pi\sum_{2,3,4} \lam_{1234}^2\prod_{i=1}^4 n_i\, \Big(\frac{1}{n_1} {+} \frac{1}{n_2}{-}\frac{1}{n_3} {-} \frac{1}{n_4}\Big) \delta(\o_{p_3p_4; p_1p_2})\frac{1}{|1-\mL_1(p_1, p_2) |^2} 
\ee

\subsection{Sum-factorized couplings}\label{sec:add}
We now  solve (\ref{SDeqn}) for a different kind of  bare coupling, which splits into a sum:
\be \label{B34a}
\lam_{1234} = \lam_{12} + \lam_{34}\,.
\ee
Here $\lam_{12}$ only depends on $p_1$ and $p_2$, and $\lam_{34}$ only depends on $p_3$ and $p_4$. Of course, there is a momentum conserving delta function, so these momenta aren't entirely independent. We make the ansatz that the sum is,	
\be  \label{B35}	
L(1,2,3,4) =- 4 i	
 \prod_{i=1}^4 D_i \, 2\pi \delta(\o_{12;34})	
\Big\{ \lam_{12}\lam_{34} X + \(\lam_{12} + \lam_{34}\) Y_s+ \(\lam_{12}y' - \lam_{34}y'^*\)(g_1^* {+} g_2^*) ( g_3 {+} g_4)  + Z\Big\}	
\ee	
where	
\bea \nn	
X &=& x(g_1^* {+} g_2^*)- x^*( g_3 {+} g_4) 	
+ x'(g_1^* {+} g_2^*) ( g_3 {+} g_4)\\ \nn 	
Y_s &=& y(g_1^* {+} g_2^*)- y^*( g_3 {+} g_4)\\	
Z &=& z(g_1^* {+} g_2^*)- z^*( g_3 {+} g_4)	
+ z'(g_1^* {+} g_2^*) ( g_3 {+} g_4)~. \label{B36}	
\eea	
Our form of the ansatz is chosen to reflect the symmetry of the four-point function, $L(1,2,3,4) = L^*(3,4,1,2)$. The coefficients that we need to determine are $x,y,z, x', y', z'$. 	
We now insert this ansatz into the integral equation (\ref{SDeqn}) governing the sum of bubble diagrams. We first evaluate the frequency integrals on the right-hand side of (\ref{SDeqn}), by making use of (\ref{OneloopInt}), followed by the momentum integrals. Note that when evaluating the momentum  integrals we make the assumption that  $x,y,z,x',y',z'$  are independent of $p_5, p_6$. For the one-loop result we see this explicitly, since the  only dependence of $N(5,6)$ is on $\o_3$ and $\o_4$, and not on e.g. $\o_{p_3}$ and $\o_{p_4}$. We may therefore move $x,y,z,x',y',z'$ outside of the integral over $p_5$. Our result  will validate that this was a self-consistent assumption.
We define, for index $r=0,1,2$,
\be
N_r = -2i \sum_{p_5} \lam_{56}^r N(5,6)~, \ \ \ \ M_r =  -2i \sum_{p_5} \lam_{56}^r M(5,6)~,
\ee
where $N(5,6)$ and $M(5,6)$ were defined in (\ref{NM}). Upon
equating  the left and right hand sides of the Schwinger-Dyson equation (\ref{SDeqn}),  we match the terms multiplying $(g_1{+}g_2)$ to find that
\be
x = x N_{1} + y N_{ 0}~, \ \ \ \
y-1 = x N_{2} + y N_{ 1}~,\ \ \
y-1 = z N_{0} + y N_{ 1}~, \ \ \
z = z N_{1} + y N_{ 2}~,
\ee
where these equations come from matching the $\lam_{12} \lam_{34}, \lam_{12}, \lam_{34}$ and coupling-independent terms, respectively.
Solving gives,
\be \label{B38}
x = \frac{N_{0}}{(N_{1} {-} 1)^2 - N_{ 0} N_{2}}~, \ \ \ y= \frac{1-N_{1}}{(N_{1} {-} 1)^2 - N_{ 0} N_{2}}~, \ \ \ z= \frac{N_{2}}{(N_{1} {-} 1)^2 - N_{ 0} N_{2}}~.
\ee
Now, matching the $(g_1^*{+}g_2^*) (g_3{+}g_4)$ terms on the  left and right hand sides of the Schwinger-Dyson equation gives,
\bea \nn
x' &=& x' N_{1}^* + y' N_{ 0}^* + x M_{1} + y M_{ 0}~, \ \ \ \
y'= x' N_{2}^* + y' N_{ 1}^* +x M_{2} + y M_{1} \\
-y'^* &=& z' N_{0}^* - y'^* N_{ 1}^* + z M_{0} + y M_{1}~, \ \ \ \
z' = z' N_{1}^* - y'^* N_{ 2}^* +z M_{1} + y M_{2}~.
\eea
The solution is
\bea \nn
x' &=& \frac{M_0 |N_1 {-} 1|^2 +  M_2 |N_0|^2  + M_1(N_0 + N_0^* - N_0 N_1^* -N_0^* N_1)}{|(N_{1} {-} 1)^2 - N_{ 0} N_{2}|^2}\\ \nn
y' &=& \frac{M_0 N_2^*(1{-}N_1) + M_2 N_0(1{-}N_1^*) + M_1(1{-} N_1 {-}N_1^* {+} |N_1|^2 {+} N_0 N_2^*)}{|(N_{1} {-} 1)^2 - N_{ 0} N_{2}|^2}\\
z' &=& \frac{M_0 |N_2|^2 +  M_2 |N_1 {-} 1|^2 + M_1\(N_2(1{-}N_1^*) + N_2^*(1{-}N_1)\)}{|(N_{1} {-} 1)^2 - N_{ 0} N_{2}|^2}~. \label{B40}
\eea
Expanding to lowest order in  $\lam$, we recover the tree-level  answer $y=1$ and everything else is zero. Expanding to next order we get the sum of the tree level and one-loop terms, $ x= N_0 $, $y=1 + N_{1}$, $z=N_2$, $x'= M_0$, $y'= M_1$, $z' = M_2$.

We now Fourier transform to get that the contribution to the kinetic equation is
 \bml \label{sumFact4}
 \frac{\d n_1}{\d t} =  4 \prod_{i=1}^4 n_i\, \frac{1}{\o_{p_3p_4; p_1p_2} {+} i\eps}\Big\{ \(\frac{1}{n_1}{+}\frac{1}{n_2}\)\frac{\lam_{12} \lam_{34} \mL_0 + \lam_{1234} (1-\mL_1) + \mL_2}{(1- \mL_1)^2 - \mL_0 \mL_2}  - \(1,2\leftrightarrow 3,4\)^* \
  \Big\}~ \\
+ 4\prod_{i=1}^4 n_i\Big(\frac{1}{n_1}{+}\frac{1}{n_2}\Big)\Big(\frac{1}{n_3}{+}\frac{1}{n_4}\Big)\,  \sum_{p_5}\frac{2n_5 n_6 }{(\o_{p_5p_6; p_1 p_2} {+} i\eps)(\o_{p_5 p_6; p_3p_4} {-} i\eps)}  \frac{\lam_{12} \lam_{34} \mathbf{x}' {+} \lam_{12} \mathbf{y}' {-} \lam_{34}\mathbf{y}'^*{+} \mathbf{z}'}{|(1{-} \mL_1)^2 - \mL_0 \mL_2 |^2}
\end{multline}
where all the $\mL_r$ in the first term are the $\mL_r(p_3, p_4)$, where
 \be \label{mLrapp}
\mL_r(p_3, p_4)= \sum_{p_5} \lam_{56}^r \frac{2 (n_5 {+} n_6)}{\o_{p_3p_4;p_5p_6} {+} i \eps}~.
 \ee
and all the $\mL_r$ in the second line of (\ref{sumFact4})  are $\mL_r(p_5, p_6)$ and
\bea \nn 
\mathbf{x}' &=& |\mL_1 {-} 1|^2 +  \lam_{56}^2 |\mL_0|^2  +\lam_{56}(\mL_0 + \mL_0^* - \mL_0 \mL_1^* -\mL_0^* \mL_1)\\ \nn
\mathbf{y}' &=& \mL_2^*(1{-}\mL_1) + \lam_{56}^2 \mL_0(1{-}\mL_1^*) + \lam_{56}(1{-} \mL_1 {-}\mL_1^* {+} |\mL_1|^2 {+} \mL_0 \mL_2^*)\\
\mathbf{z}' &=& |\mL_2|^2 +  \lam_{56}^2 |\mL_1 {-} 1|^2 + \lam_{56}\(\mL_2(1{-}\mL_1^*) + \mL_2^*(1{-}\mL_1)\)~.
\eea

\subsection{Optical and spin-wave turbulence}\label{sec:diff}

Optical and spin-wave turbulence correspond to the case of sum factorized couplings. Let us discuss this case a bit more, on the basis of the ``renormalized'' kinetic equation found from summing bubble diagrams, (\ref{sumFact4}).  The most dramatic difference from the kinetic equation for the product-factorized vertex (\ref{423}) is that $\mL_2$ generally diverges in the UV: $\mL_{ 2}\simeq -\lambda^2k_{max}^{2\beta+d-\gamma-\alpha}=-\lambda^2k_{max}^{4\beta/3-\alpha}$. Note also that $\Re \mL_0$ always diverges in the IR, since $\gamma{=}d{+}2\beta/3{>}d$,   so that  $\Re\mL_{ 0}(p_1,p_2)\approx  3k_0^{-2\beta/3}/2\beta[{\omega_{{\bf p}_1}{+} \omega_{{\bf p}_2}{-}
\omega_{{\bf p}_1+{\bf p}_2} }]$. This divergence is stronger than that of $\mL_1$. That means that we can neglect the term  $\mL_1\lambda_{1234}$ in the nominator, which is always smaller than $\mL_0\lambda_{12}\lambda_{34}$.

When all the  $\mL_{i}$ are small, there two types of corrections to the weak-turbulence solution, one enhanced by an IR divergence, and the other by  a UV divergence:
\bea&{\delta n_k\over n_k}\simeq \lambda_{1234} \mL_{0}\simeq\epsilon_k\left({k\over k_0}\right)^{2\beta/3}=\epsilon_0\left({k\over k_0}\right)^{\beta-\alpha}\,,\nonumber\\& {\delta n_k\over n_k}\simeq {\mL_{ 2}\over\lambda_{1234}}\simeq\epsilon_k \left({k_{max}\over k}\right)^{4\beta/3-\alpha} \,.\label{NL2}\eea
We see that the main prediction about enhancement of non-linearity by non-locality holds. Distortion enhanced by IR non-locality  grows faster (or decays slower) for (\ref{NL2}) than for  (\ref{var}). On the contrary, the new  UV- enhanced distortion  decays with $k$.

The dimensionless couplings that determine the denominator are  $\mL_1$ and  $\mL_0\mL_2$.  Since we now have $\mL_2$ growing with $k_{max}$ for a weak-turbulence solution, the limit $k_{max}\to\infty$ at finite $k,k_0$ is non-trivial, in distinction from the product factorized coupling case. Similar to the transition to  strong turbulence, discussed in the previous section, we see two possible scenarios, depending on whether the denominator of the renormalized vertex, $(1{-} \mL_1)^2{ -} \mL_0 \mL_2$,  can or cannot  approach zero. Since we always have $\Re \mL_2<0$,  the sign of $\mL_0\mL_2$ is determined by the dispersion relation: ${\rm sign}\,\Re \mL_{ 0}={\rm sign} (\omega_{\bf p_1}{+} \omega_{\bf p_2}{-}
\omega_{{\bf p}_1{+}{\bf p}_2})={\rm sign}\,(1{-}\alpha) $. For $\alpha>1$, confinement (turbulence dominated by bound states) may appear when $(1-\mL_1)^2\approx \mL_0\mL_2$. On the other hand, for $\alpha<1$, one can imagine  strong-turbulence solutions with large $\mL_2$. In this case, the effective interaction behaves as $T_{1234}\rightarrow-1/\mL_{0}=-1 k_0^{2\beta/3}k^{\alpha}$, which , remarkably, is independent of the bare vertex.  Such turbulence is again independent of $k_{max}$, yet the bare kinetic equation  is replaced by a strongly renormalized one. This is purely hypothetical at this stage; the analysis of strongly renormalized cases is left for the future.

The terms at the  next order in $\lambda$ are proportional to $(n_1^{-1}+n_2^{-1})(n_3^{-1}+n_4^{-1})$, see (\ref{B40});   they produce a UV divergence, cut off by the denominators. The most divergent term in the numerator (after the Fourier transform) is $M_2 N_1$, which gives $\delta n_k/n_k\simeq \epsilon_k (k_{max}/k)^{5\beta/6-\alpha-2\kappa}$. It is restricted when $\mL_{1}(k,k_{max})\simeq \epsilon_k(k_{max}/k)^{\beta/2-\kappa}(k/k_0)^{\beta/6}\simeq 1$, which makes these terms proportional to non-integer powers of $\lambda$, like in the product factorized case.

Let us briefly discuss a few physically important cases. The simplest is the nonlinear the so-called optical turbulence described by Schr\"odinger equation (referred to as the Gross-Pitaevskii  equation when applied to cold atoms or bosons). This is  a universal model that describes the nonlinear dynamics of a spectrally narrow distribution, as well as many other systems. It corresponds $\alpha=2$, $d=3$ and $\beta=0$. This also corresponds to Langmuir wave turbulence in non-isothermal plasmas (hot electrons, cold ions). The direct energy cascade $n_k\propto k^{-3}$ gives a  collision integral which converges in the UV, while being on the verge of diverging in the IR. Numerical solutions of the kinetic equation show that the spectrum is realized with a weak (logarithmic) distortion \cite{FR}. Our considerations show that all the high-order terms converge, so  one can use the kinetic equation for arbitrarily long direct cascades.

Spin waves with exchange interaction  correspond to $\alpha=\beta=2$ and a sum factorization of the bare vertex, $\lambda_{1234} =-\lambda[(\vec p_1\cdot\vec p_2)+(\vec p_3\cdot\vec p_4)]$ \cite{ZLF,spin1,spin2}, so that the naive dimensionless nonlinearity parameter decays along the cascade as $\epsilon_k=\epsilon_0(k_0/k)^{4/3}= \lambda k^{-7/3}$. Our computation gives $\mL_{0}=-3k_0^{-4/3}/4k_1k_2$ and $\mL_{2}=-3\lambda^2k_{max}^{2/3}/2d$, while $\mL_{1}\simeq \lambda k^{-7/3}\simeq \epsilon_k$ is given by a convergent integral. The small corrections (\ref{NL2}) both have the same (positive) sign, since  renormalization decreases the interaction and thus increases the turbulence level. How weak turbulence becomes strong upon  increase of $k_{max}$ is determined by the dimensionless parameter $\mL_{2}/\lambda_{1234} \simeq \epsilon_0 (k_{max}/k_0)^{2/3}(k_0/k)^2$, that is, by the competition between the pumping-set nonlinearity level $\epsilon_0$  and the length of the cascade $k_{max}/k_0$ (analog of the Reynolds number).
If at $k\simeq k_0$ we have  $ \epsilon_0(k_{max}/k_0)^{2/3}<1$, then  (\ref{NL2}) shows that the effective nonlinearity is $\epsilon_0$, which is independent of $k$. If, however, the cascade is long enough, so that $ \epsilon_0(k_{max}/k_0)^{2/3}>1$, we have strong turbulence already at the pumping scale, even though $\epsilon_0$ is small. When the effective nonlinearity is not small, it is the denominator in (\ref{423}) which determines the renormalization. That reverses the trend and increases the interaction,
since $\mL_{1}>0$ and  $\mL_{0}\mL_{2}>0$. We see that in this case, in distinction from the product-factorized case determined by the single loop integral $\mL_1$, it is not enough to consider the one-loop approximation to determine whether four-wave scattering is enhanced or suppressed by multi-wave interactions. Whether the denominator in the renormalized vertex,  $(1{-}\mL_1)^2{ -} \mL_0 \mL_2$,  could approach zero and we could have turbulence dominated by bound states of spin waves deserves further study. A  direct cascade in an isothermal plasma is expected to behave similarly and will also be analyzed elsewhere. A detailed application of our approach to  plasma turbulence could be important  for thermonuclear studies where it may be necessary to go outside of the weak turbulence approximation even for weak nonlinearities, despite the current belief to the contrary.

We see that even though the exact form of the renormalized kinetic equation depends on the vertex, its structure and the basic properties of the corrections to the weak-turbulence solutions are determined by the asymptotics of the loop integrals.   We may then make plausibe assumptions about the physical cases, where $\lambda_{1234}$ does not have a simple form.
For a generic $\lambda_{1234}$, which neither factorizes into a product nor breaks up into sum, the solution of the integral equation giving the sum of bubble diagrams, see (\ref{SDeqn}), cannot be generally written via a geometric series.
 If, however, the effects we are interested in (like non-analytic corrections due to the UV cutoff and the transition to  strong turbulence) are determined by the asymptotics with $p_2,p_4,p_6\ldots\gg p_1,p_3,p_5\ldots$, then the vertex factorizes according to  (\ref{beta_1}). That means that even though (\ref{RKE})  is no longer an exact solution of the integral equation (\ref{SDeqn}), the  integral of its Fourier transform over $p_2,p_3,p_4$ (that is the renormalized kinetic equation) can be used to describe such effects. Using at  $p_2,p_4\gg p_1,p_3$ the general asymptotics (\ref{beta_1}), $\lambda_{1234}=\lambda(p_2p_4)^{\beta_1/2}(p_1p_3)^{\beta-\beta_1)/2}$, we estimate the effective UV cutoff as $p_{max}\propto \lambda^{1/(\kappa-\beta)}$. We can also estimate $\Re \mL_1\simeq \lambda k_0^{\beta/3-\beta_1}k^{\beta_1}/\tilde\omega_k$. It has an IR divergence when $\beta_1{>}\beta/3$, exactly like for frequency renormalization. When $\Re \mL_1\simeq \lambda k_0^{\beta/3-\beta_1}k^{\beta_1}/ \omega_k\propto k^{\beta_1-\alpha}$ grows along the cascade, it approaches unity right where the frequency renormalization is getting substantial; at larger $k$, we obtain saturation: $\Re \mL_1\simeq \lambda k_0^{\beta/4-\beta_1}k^{\beta_1}/\tilde\omega_k\simeq 1$.

\subsection{General coupling} \label{sec:appB4}
Finally, let us look at the case of general couplings. If we want the contribution to the kinetic equation of the full sum of all bubble diagrams then there is no choice but to solve the integral equation (\ref{SDeqn}), which is written in frequency space, and then take its Fourier transform. However, we have seen that to understand how the one-loop UV divergences are cured we only need the $(n_1^{-1}+n_2^{-1})$ and $(n_3^{-1}+n_4^{-1})$ terms in the four-point function (equivalently, kinetic equation), and not the $(n_1^{-1}+n_2^{-1}) (n_3^{-1}+n_4^{-1})$ terms. The former all come from monotonic time orderings. To see this, we notice that the only way to avoid a term $(n_1^{-1}+n_2^{-1}) (n_3^{-1}+n_4^{-1})$ is for either the time at the vertex directly after $t_a$ when going clockwise ($t_b$) to be the latest, or for the time at the vertex directly before $t_a$ ($t_d$, in the two loop case in Fig.~\ref{P4loop}) to be the latest. Because neighboring vertices must have neighboring times, in order for the diagram to give a nonvanishing contribution, we conclude that the time orderings from the vertex directly after $t_a$ to directly before $t_a$ must all be either increasing or decreasing.

\begin{figure}[h]
\centering
\includegraphics[width=2.8in]{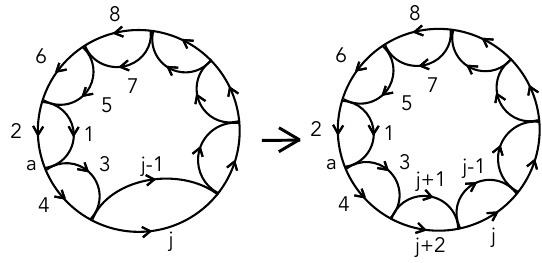}
\caption{}\label{Pmanyloopapp2}
\end{figure}

Consider the term with $(n_1^{-1}+n_2^{-1})$. Let us take a bubble diagram contributing to the four-point function that has $j/2{-}1$ loops. Now add one more loop, which we place at the end, see Fig.~\ref{Pmanyloopapp2}.  For this monotonic time ordering, for the term in the kinetic equation with $(n_1^{-1}+n_2^{-1})$, the addition of one more loop causes us to replace,
\be
\lam_{j{-}1, j, 3,4} \rightarrow \sum_{p_{j{+}1}, p_{j{+}2}}\!\!\!\lam_{j{-}1,j, j{+}1, j{+}2} \lam_{j{+}1, j{+}2, 3,4}n_{j{+}1}n_{j{+}2} \Big(\frac{1}{n_{j{+}1}} +\frac{1}{n_{j{+}2}}\Big) \frac{1}{\o_{p_3p_4;p_{j{+}1},p_{j{+}2}}{+}i\eps}~.
\ee
Therefore, the contribution to the equal-time four-point function coming from the bubble diagram which contains  $(n_1^{-1}+n_2^{-1})$, as well as the one that contains  $(n_3^{-1}+n_4^{-1})$, is,
\be
 \prod_{i=1}^4 n_i\, \frac{1}{\o_{p_3p_4; p_1p_2} {+} i\eps}\Big( \Big(\frac{1}{n_1} {+} \frac{1}{n_2}\Big) T(p_1, p_2, p_3, p_4) - \Big(\frac{1}{n_3} {+} \frac{1}{n_4}\Big) T(p_3, p_4, p_1, p_2)^*\Big)~,
\ee
where $T(p_1, p_2, p_3, p_4)$ is the solution to the integral equation,
\be \label{Teqn}
T(p_1, p_2, p_5, p_6) = \lam_{1256} + \sum_{p_7,p_8} T(p_1,p_2, p_7, p_8) \lam_{7856} l(p_7,p_8)~,  \ \ \ \ l(p_7, p_8) =  \frac{2(n_7 {+}n_8)}{\o_{p_3p_4;p_7p_8}{+}i\eps}~.
\ee
It is straightforward to check that for couplings that have product or sum factorization we recover the  solutions given earlier.

\end{document}